\documentclass{aa}

\usepackage{graphicx}
\usepackage[varg]{txfonts}
\usepackage{natbib,twoopt}
\usepackage[breaklinks=true]{hyperref} 
\bibpunct{(}{)}{;}{a}{}{,} 
\makeatletter
\newcommandtwoopt{\citeads}[3][][]{\href{http://adsabs.harvard.edu/abs/#3}%
{\def\hyper@linkstart##1##2{}%
\let\hyper@linkend\@empty\citealp[#1][#2]{#3}}}
\newcommandtwoopt{\citepads}[3][][]{\href{http://adsabs.harvard.edu/abs/#3}%
{\def\hyper@linkstart##1##2{}%
\let\hyper@linkend\@empty\citep[#1][#2]{#3}}}
\newcommandtwoopt{\citetads}[3][][]{\href{http://adsabs.harvard.edu/abs/#3}%
{\def\hyper@linkstart##1##2{}%
\let\hyper@linkend\@empty\citet[#1][#2]{#3}}}
\newcommandtwoopt{\citeyearads}[3][][]%
{\href{http://adsabs.harvard.edu/abs/#3}
{\def\hyper@linkstart##1##2{}%
\let\hyper@linkend\@empty\citeyear[#1][#2]{#3}}}
\makeatother

\begin{document}

\title{Fragmentation of electric currents in the solar corona by plasma flows}

\author{D.H. Nickeler\inst{1}, M. Karlick\'y\inst{1},
T. Wiegelmann\inst{2} \and M. Kraus\inst{1}
}

\institute{Astronomical Institute, AV \v{C}R, Fri\v{c}ova 298,
25165 Ond\v{r}ejov, Czech Republic\\
\email{dieter.nickeler@asu.cas.cz}
\and
Max-Planck Institute for Solar System Research, Katlenburg-Lindau, Max-Planck Strasse 2,
37191 Katlenburg-Lindau, Germany
}

\date{Received; accepted}

\authorrunning{Nickeler et al.}

\abstract {} 
{We consider a magnetic configuration consisting of an arcade structure and a detached 
plasmoid, resulting from a magnetic reconnection process, as is typically found in 
connection with solar flares. We study spontaneous current fragmentation caused by shear 
and vortex plasma flows.} 
{An exact analytical transformation method was applied to calculate self-consistent
solutions of the nonlinear stationary magnetohydrodynamic equations. The
assumption of incompressible field-aligned flows implies that both the
Alfv\'{e}n Mach number and the mass density are constant on field lines. We
first calculated nonlinear magnetohydrostatic equilibria with the help of the
Liouville method, emulating the scenario of a solar eruptive flare
configuration with plasmoids (magnetic ropes or current-carrying loops in 3D)
and flare arcade. Then a Mach number profile was constructed that describes the
upflow along the open magnetic field lines and implements a vortex flow inside
the plasmoid. This Mach number profile was used to map the magnetohydrostatic
equilibrium to the stationary one.} 
{We find that current fragmentation takes
place at different locations within our configuration. Steep gradients of the
Alfv\'{e}n Mach number are required, implying the strong influence of shear
flows on current amplification and filamentation of the magnetohydrostatic
current sheets. Crescent- or ring-like structures appear along the outer
separatrix, butterfly structures between the upper and lower plasmoids, and
strong current peaks close the lower boundary (photosphere). Furthermore,
impressing an intrinsic small-scale structure on the upper plasmoid results in
strong fragmentation of the plasmoid. Hence fragmentation of current sheets and
plasmoids is an inherent property of magnetohydrodynamic theory.}
{Transformations from magnetohydrostatic into magnetohydrodynamic steady-states
deliver fine-structures needed for plasma heating and acceleration of particles
and bulk plasma flows in dissipative events that are typically connected to magnetic
reconnection processes in flares and coronal mass ejections.}

\keywords{Magnetohydrodynamics (MHD) -- Sun: flares -- Sun: corona -- 
methods: analytical}

\maketitle

\section{Introduction}

In the standard flare scenario \citepads[e.g.,][]{1996ApJ...466.1054M} the
energy release of the primary flare (primary magnetic reconnection) takes place in the
current sheet below a rising magnetic rope. Here, the plasmoids (the secondary
magnetic ropes in 3D), which are a natural outcome of the reconnection process,
are formed and ejected. The ejection of plasmoids can be traced observationally
via soft X-ray and radio waves, which map the magnetic-field reconnection
\citepads{1998ApJ...499..934O, 2000A&A...360..715K, 2002A&A...395..677K, 2004A&A...417..325K}. 
With increasing spatial resolution of the solar photosphere and chromosphere,
flares, jets, and plasmoids on different scales are observed
\citepads[e.g.,][]{2013Natur.493..485C, 2013Natur.493..501C}. This means that
the solar atmosphere is highly structured, and magnetic reconnection processes
are ubiquitous. As such, the current sheets initiating reconnection processes
cannot be smooth but must contain some internal fragmented structure.
Consequently, magnetic reconnection itself must be fractal. As an efficient
mechanism to cascade down to smaller scales, instabilities have proven to be an
ideal trigger.

Many nonlinear, time-dependent magnetohydrodynamic (MHD) simulations focus on linear 
and nonlinear instabilities, which are initiated via arbitrarily prescribed small 
perturbations
of an initially smooth, static equilibrium. These instabilities typically result in
reconnection, and in the following in fragmentation of the magnetic field and hence
the current density \citepads[e.g.,][]{2010AdSpR..45...10B, 2011ApJ...733..107K,
2011ApJ...737...24B}, forming chains of plasmoids \citepads{2007PhPl...14j0703L,
2010PhRvL.105w5002U}, coalescence and further fragmentation of plasmoids
\citepads{2010AIPC.1242...89P, 2011ApJ...733..107K}, and plasmoids on progressively smaller
scales \citepads{2001EP&S...53..473S}.

The process of cascading can also be initiated by stochastic velocity
fluctuations, generating small-scale structures of the large-scale magnetic
field \citepads{1999ApJ...517..700L, 2009ApJ...700...63K, 2011PhRvE..83e6405E}.
This turbulent approach, however, originates from external perturbations
impressed on initial background (magnetic and velocity) fields, requiring the
prescription of initial noise, e.g., in the form of power-law spectra of
perturbations. On the other hand, the turbulent reconnection can result
from a successive coalescence and fragmentation of plasmoids, their fast
heating, and an increase of the plasma beta parameter at some locations, where
the flow instabilities become important as well \citepads{2012A&A...541A..86K}.

In contrast to studies using instabilities or turbulence as the initial trigger for 
fragmentation, the MHD theory itself inherently provides the cradles for fractal 
structures, because the MHD is scale-free and therefore applies to large as well as to 
small scales \citepads{2012EGUGA..14.3226S, 2012cosp...39.1785S}. In their studies 
of the Earth's magnetotail using a quasi-static adiabatic MHD approach, 
\citetads{1995GeoRL..22.2057W} previously noticed the fragmentation of a
thin current sheet. In their investigations, they found the formation of
double-structures of the current density when using nonsimilarity solutions of the
quasi-static equations. Similarly, the numerical investigations of 
\citetads{2001JGR...106.3811B} revealed the formation of thin current sheets from a 
sequence of static equilibria. Thus, instead of using perturbations of a smooth,
static equilibrium, one might start directly from already structured, fragmented MHD
equilibrium states. For this, one needs to construct a selfconsistent analytical 
description of the time-independent, nonlinear dynamics
\citepads[see, e.g.,][]{2001ohnf.conf...57N, 2006A&A...454..797N, 2010AnGeo..28.1523N,
2012AnGeo..30..545N}.

Separatrices form during magnetic reconnection processes, which originate in
so-called X-points. These X-points can separate regions of closed and open field
lines. The open field-line regions can be regarded as field lines along which, e.g.,
the solar wind can flow into the interplanetary space, while the closed regions correspond
to, e.g., magnetic arcades or flux ropes (plasmoids) from which plasma cannot leave. To
stabilize such a configuration, in which strong flows occur outside and (almost) no flows
inside, shear currents have to keep the system in equilibrium. Hence the physical problem 
can be approximately described with the static approach in regions of closed field lines,
while in regions with open field lines the problem is in steady-state \citepads[see,
e.g.,][]{2006A&A...454..797N, 2010AnGeo..28.1523N, 2012AnGeo..30..545N}.

\citetads{2010AnGeo..28.1523N} considered that the
Alfv\'en Mach number, $M_{A}$, determining the strength of the flow and therefore the
plasma velocity, vanishes within the plasmoid, so that the structure is basically
magneto-hydrostatic (MHS). However, not every closed
field-line region must necessarily be of MHS nature. Instead, plasmoids might contain
vortices, because slight asymmetries during the ejection event could result in a
nonzero angular momentum transfer. Hence the flow inside plasmoids can be sheared.
Shear flows were found to produce current filamentation not only in solar or magnetospheric
environments, but also in space and astrophysical
plasmas, e.g., in astrophysical jets, where shear flows also induce the filamentation of
currents \citepads{1998PhPl....5.3732W, 2000PhPl....7.5159K}.

In this paper, we investigate the role of shear flows within a configuration containing 
a magnetic dome and detached plasmoids, resembling a typical solar-flare configuration after
a first reconnection process. In our investigations, we used a selfconsistent analytical
description of the time-independent, nonlinear dynamics. The paper is
structured as follows: in Sect.\,\ref{sec2} we introduce the basic equations
and the transformation method, while the results are described in
Sect.\,\ref{sec3}. The assumptions are discussed in Sect.\,\ref{sec4}, and the
conclusions are given in Sect.\,\ref{sec5}.

\section{Basic equations}
\label{sec2}

 \begin{figure}
   \centering
   \includegraphics[width=0.6\hsize]{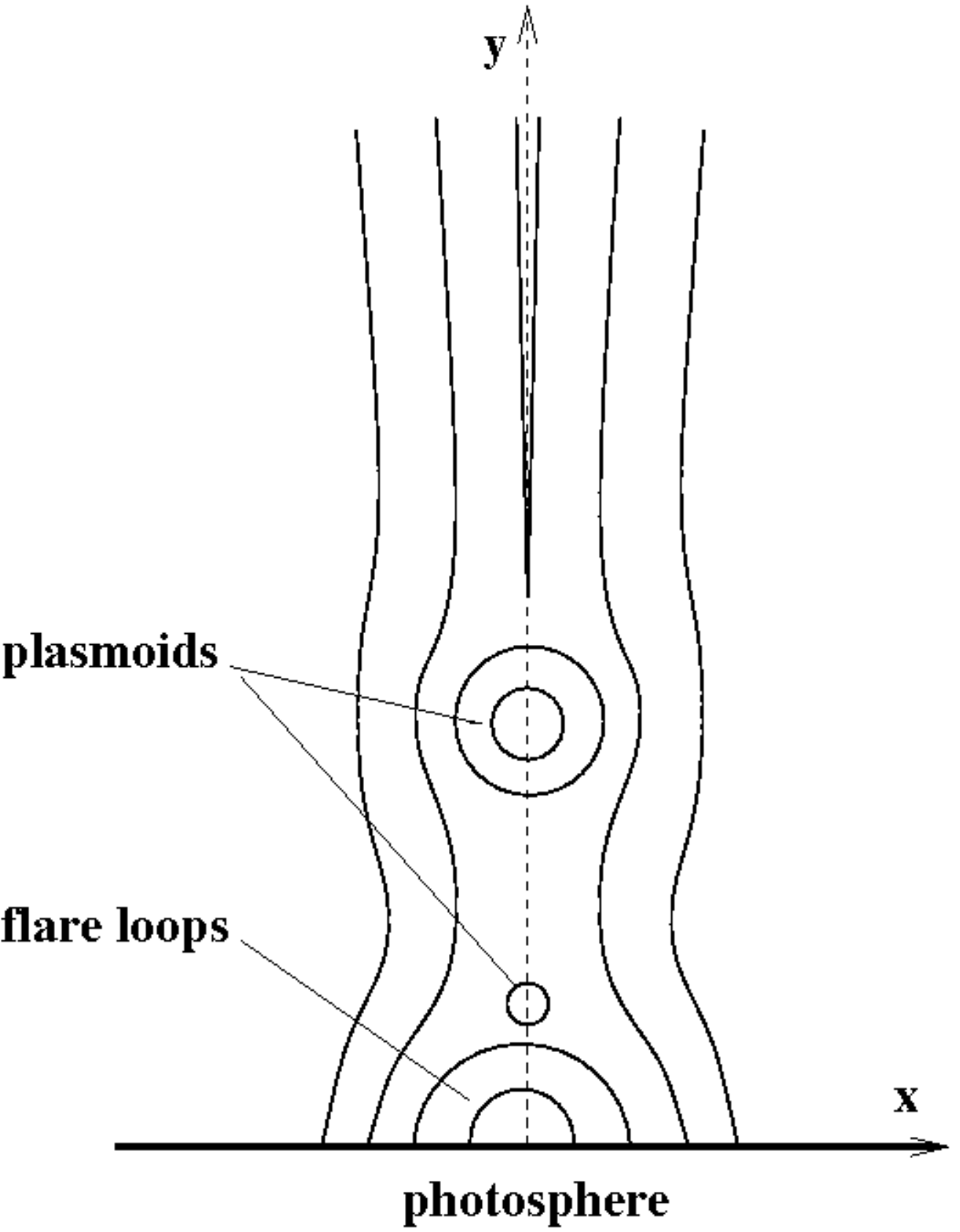}
      \caption{Sketch of a magnetic configuration of a solar flare with plasmoids
      formed via magnetic reconnection processes.}
         \label{sketch}
   \end{figure}

We assumed a magnetic configuration of a solar flare with a plasmoid formed via 
some first magnetic reconnection event (see Fig.\,\ref{sketch}). The plasmoid is 
enclosed by two X-points, while the plasmoid itself hosts a magnetic O-point in its center.
The equilibrium of such a system can be formulated using the
steady/static approach. This is, however, only strictly valid if the plasmoid
has no or only marginal motion in $y$ direction. The dynamics of plasmoids
depends on the reconnection rate at the X-points below and above the plasmoid,
and plasmoids ``in rest'' \citepads[see, e.g.,][] {2008SoPh..253..173B,
2008A&A...477..649B} and stable \citepads[i.e. without further coalescence,
see, e.g.,][]{2006PhPl...13c2307K} have been found by numerical simulations,
justifying our assumption.

The choice of a certain equilibrium defines an arbitrary, but fixed length-scale. 
This does not allow one to make inferences on the properties of the plasmoid on (much) 
smaller scales, on which, e.g., stationary shear flows related to vortex sheets might 
exist. Such shear flows would generate additional forces on the former MHS states, 
which can only be compensated for by changes in Lorentz forces and
pressure gradients. To maintain the force balance self-consistently, we applied
the transformation method developed by \citetads{1992PhFlB...4.1689G} and
advanced/progressed by
\citetads{1999GApFD..91..269P} and \citetads{2006A&A...454..797N}. In the past
decades many attempts have been made to find exact and analytical solutions of nonlinear
steady-state (= stationary) MHD equations \citepads[e.g.,][]{1981ApJ...245..764T,
1996ApJ...460..185C, 1997PhPl....4.3544G, 2005AdSpR..35.2067N, 2006AdSpR..37.1292N}.
However, the transformation is the only systematic method that physically and mathematically
relates steady-state MHD flows to MHS states. For such a transformation
to work, it is reasonable to request that in the stationary state
the velocity field and the magnetic field are parallel (field-aligned flows).
This guarantees that the electric field vanishes, according to the ideal Ohm's law
\begin{equation}
\vec E + \vec v \times \vec B = \vec 0 \quad
\Rightarrow \quad \vec E = \vec 0\, .
\end{equation}
We note that other transformations between steady MHD states exist, which lead to 
configurations in which the velocity field and the magnetic field are not necessarily parallel
\citepads{2000PhRvE..62.8616B, 2001PhLA..291..256B, 2002PhRvE..66e6410B}. However, only in
the case of incompressible field-aligned flows one can always reduce the steady-state MHD
equations to the MHS equations. Another advantage of the transformation method is that it
is independent of the dimensions, i.e., it can be performed in 1, 2, and 3D.

\subsection{Transformation from MHS states to stationary MHD configurations}
\label{trafo}

In the following we restrict the analysis to sub-Alfv\'enic flows to emphasize
in particular their relationship to MHS states. In addition, we use normalized
parameters, for which we introduce normalization constants $\hat{B}, \hat{\rho}, \hat{l}, 
\hat{p}$ and $\hat{\varv}_{A}$, where $\hat{\varv}_{A}=\hat{B}/\sqrt{\mu_{0}\hat{\rho}}$
is the normalized Alfv\'en velocity. Let $\vec v$ be the plasma velocity normalized on 
$\hat{\varv}_{A}$, $\rho$ the mass density normalized on $\hat{\rho}$,
$\vec j=\vec\nabla\times\vec B$ the current density vector normalized on
$\hat{B}/(\mu_{0}\hat{l})$ with $\hat{l}$ as the characteristic length scale, and
$p$ the scalar plasma pressure normalized on
$\hat{p}=\hat{B}^2/\mu_{0}$.
With these definitions, we can write the set of equations of stationary, field-aligned
incompressible MHD, consisting of the mass continuity equation, the Euler equation, the
definition for field-aligned flow and Alfv\'{e}n Mach number, the incompressibility 
condition, and the solenoidal condition for the magnetic field, in the form
\begin{eqnarray}
\vec\nabla\cdot(\rho\vec v) & = & 0\, ,\label{konti}\\
\rho\left(\vec v\cdot\vec\nabla\right)\vec v & = & \vec j\times\vec B - \vec\nabla p\, ,
\label{euler0} \\
\vec v & = & \frac{M_{A}\vec B}{\sqrt{\rho}}\, ,\label{paral}\\
\vec\nabla\cdot\vec v & =& 0\, ,\label{konti2}\\
\vec\nabla\cdot\vec B & = & 0\label{divb} \, .
\end{eqnarray}

This set of equations can always be reduced to the set of static equations using
the transformation equations \citepads[for details see][]{2010AnGeo..28.1523N,
2012AnGeo..30..545N} of the form

\begin{eqnarray}
\vec B &&=\frac{\vec B_{S}}{\sqrt{1-M_{A}^2}}\, ,\label{magtrafo} \\
p &&=p_{S} - \frac{M_{A}^2\left|\vec B_{S}\right|^2}{1-M_{A}^2}
\, , \label{pressuretrafo}\\
\sqrt{\rho} \vec v &&=
\frac{M_{A}\vec B_{S}}{\sqrt{1-M_{A}^2}}
\equiv\ M_{A}\vec B\label{streaming2} \, , \\
\vec j &&=\frac{M_{A}\,\vec\nabla M_{A}\times\vec B_{S}}{\left(1-M_{A}^2
\right)^{\frac{3}{2}}}
+\frac{\vec j_{S}}{\left(1-M_{A}^2\right)^{\frac{1}{2}}} \, , \label{currenttrafo}\\
\vec\nabla p_{S} &&=\vec j_{S}\times\vec B_{S} \label{mhs1}
\,\, ,
\end{eqnarray}
where the subscript $S$ refers to the original MHS fields.
Here it is a necessary condition that the Alfv\'en Mach number $M_{A}$ and the
density $\rho$ are constant along fieldlines, i.e.,
\begin{eqnarray}
\vec B\cdot\vec\nabla M_{A} & = & 0 \label{cond1} \\
\vec B\cdot\vec\nabla\rho   & = & 0\, . \label{conditions}
\end{eqnarray}
An important property of this type of transformation
is the fact that every transformed magnetic field strength $|\vec B|$ is stronger
than the original static magnetic field strength $|\vec B_{S}|$ (as long as $M_{A}\neq 0$).
Moreover, as $\vec j$ is directly proportional to the term $\vec\nabla M_{A}$, which
can have an arbitrarily (but not infinite) high value, basically every infinitesimale scale
$1/\nabla = l>0$  can be chosen. Therefore, we can produce a current that is higher than any
current threshold to excite anomalous resistivity, such that current-driven instabilities and 
hence magnetic reconnection can be induced. This generated current can be even more 
amplified when the Alfv\'en Mach number approaches the limit $M_{A}\la 1$. 
We stress that the transformation method provides a nonlinear self-consistent solution 
of the stationary MHD equations, and changing any of the physical variables produces a 
nonlinear feed-back of all other variables. Variations of $M_{A}$ should not be 
misunderstood as an explicit time-dependent change or sequence of the underlying MHS 
equilibrium, like in the quasi-static sequences of \citetads{1995GeoRL..22.2057W} or 
\citetads{2001JGR...106.3811B}. Instead, the transformation has to be interpreted as a
nonlinear variation or displacement of the former initial MHS equilibrium. That is, in
affinity to variational calculus, the steady-states are \lq located\rq~in the proximity
of MHS states.

The set of transformation equations (Eqs.\,(\ref{magtrafo} - \ref{mhs1})) together with the
conditions of Eqs.\,(\ref{cond1} - \ref{conditions}) provide a \lq recipe\rq~to construct 
field-aligned, incompressible flows along the MHS structures. In practice, we first need to 
calculate an MHS equilibrium. 
In the following we assumed that the equilibrium has some
sort of symmetry (e.g. in z-direction), so that it can be reduced to
a pure 2-dimensional (2D) problem\footnote{A restriction to pure 2D is justified, because
it enhances the clarity of the representation of the fragmentation process. Our studies
are aimed at the fragmentation of the isocontours of the current density $j_{z}$.}.
In that case, the equilibrium value of the magnetic field has the form
$\vec B_{S} = \nabla A(x,y) \times \vec e_z$.

Next, we need to determine a Mach number profile, $M_{A}(A)$. This profile has to depend 
locally only on the flux function, $A$, so that $\vec B_{S}\cdot \vec\nabla M_{A} = 0$, and 
hence the condition Eq\,(\ref{cond1}) is automatically fulfilled. 
The \lq new\rq~magnetic field, i.e., the steady-state
field, is then given by a new flux function $\alpha$, which is a function of $A$, such that
\begin{equation}
M_{A}^2=1-\displaystyle\frac{1}{\left(\frac{d\alpha}{dA}\right)^2}\quad\Leftrightarrow\quad
\left(\alpha'(A)\right)^2=\frac{1}{1-M_{A}^2} \label{Ma_alpha}
\end{equation}
and $\vec B = \vec\nabla\alpha \times \vec e_{z}$ \citepads[see][]{2010AnGeo..28.1523N}.
The prime denotes the derivative with respect to $A$. Armed with this Mach number
profile and $\vec B_{S}$, the set of transformation equations (Eq.\,(\ref{magtrafo}) to
(\ref{mhs1})) can be evaluated.

For the adopted 2D shape of the magnetic field the current-transformation equation 
(Eq.\,(\ref{currenttrafo})) takes the form
\begin{eqnarray}
j = j_{z} & = & -\frac{M_{A} M_{A}'\,\left(\vec\nabla A\right)^2}{\left(1-M_{A}^2\right)^{3/2}} -
\frac{\Delta A}{\left(1-M_{A}^2\right)^{1/2}} \\
& = & -\Delta\alpha=-\alpha'' \left(\vec\nabla A\right)^2 - \alpha' \Delta A .
\label{currenttrafoexpl6}
\end{eqnarray}
The current fragmentation is strong where the magnetic field is strong, as the increase in
current and its spatial variation is mainly governed by $M_{A}'$ but amplified by
$\left(\vec\nabla A\right)^2$.

One interesting and important property of this transformed current is the fact that
it can have a zero-crossing even for an initially monopolar MHS-current distribution.
This means that any suitable choice of a transformation can make $j_{z}$ negative
(positive), although the MHS current is completely positive (negative). In particular,
the zero-crossing of the current requires that it has to vanish at some point, i.e.,
$j = j_{z} \stackrel{!}{=} 0$. This delivers a condition for the Mach number profile
of the form
\begin{equation}
 M_{A}M_{A}'=- \frac{\Delta A}{\left(\vec\nabla A\right)^2}\,\left(1-M_{A}^2\right)\, ,
\label{determine}
\end{equation}
with the restriction $\left(\vec\nabla A\right)^2\neq 0$.
As an important example of a monopolar MHS-current we refer to Liouville's equation
given by $\Delta A=\exp(-2A)$ (see Eq.\,(\ref{lio}) below). Because $\Delta A$ is always
positive and $\left(\vec\nabla A\right)^2$ and $1-M_{A}^2$ are positive as well, the 
left-hand side derivative of Eq.\,(\ref{determine}) must be negative, i.e.,
$d/dA\,(M_{A}^2/2)<0$. This condition can in principle be fulfilled with any suitable
Alfv\'en Mach number profile that is monotonically decreasing with $A$ (at least
locally). This demonstrates the power of the transformation method and shows that it can
be used to generate strong current fragmentation.

The zero-crossing is a definite sign that fragmentation can take place, but in many cases
it is sufficient to have a strong gradient concerning $M_{A}$ and/or a large
$(\nabla A)^{2}$.
On the other hand, this means that in the vicinity of a magnetic null point $M_{A}'$ must
be extremely large to compensate the vanishing magnetic field strength.
Nevertheless, depending on the choice of the Mach number profile, current
fragmentation can happen even without a zero-crossing of the transformed current.

\section{Results}
\label{sec3}

\subsection{Nonlinear static equilibria}
\label{nonlinstat}

As described in the previous section, the first step is to derive a
reasonable initial MHS equilibrium, which is able to reproduce a field-line scenario
with individual disconnected plasmoids, as drawn schematically in Fig.\,\ref{sketch}.
For this, we used two well-known equilibrium configurations and combined them.

Starting from the static magnetic field in 2D,
$\vec{B_{S}} = \vec\nabla A \times \vec e_{z}$, and inserting it into the MHS equilibrium
equation (Eq.\,(\ref{mhs1})) delivers the well-known Grad-Shafranov-equation, often also
called L\"{u}st-Schl\"{u}ter-equation \citepads[see, e.g.,][]{1957ZNatA..12..850L,
1958JETP....6..545S}
\begin{eqnarray}
\Delta A=-\frac{dp_{S}}{dA}.
\label{GSE}
\end{eqnarray}
Because $\vec{B_{S}}\cdot \vec\nabla A = 0$ is valid, lines of constant $A$
are field lines. This implies that the current $j= -\Delta A$ is constant
along field lines, as is the pressure $p_{S}$, because they are functions of $A$,
and consequently the isocontours of the current have the same topological and geometrical
structure as those of the field lines.

For the pressure function $p_{S}(A)$ we use
\begin{eqnarray}
p_{S}(A)=\frac{1}{2}\,\exp(-2A),
\end{eqnarray}
as derived in the frame of the Vlasov theory by, e.g., \citetads{1934PhRv...45..890B},
\citetads{1962IlNC...23..115H}, and \citetads{1973JGR....78.3773K}. The same function is 
typically applied in MHS or MHD
configurations in which the pressure monotonically decreases in the direction perpendicular 
to the current sheet. Examples are flare configurations, magnetotails, and helmet streamers
\citepads[see, e.g.,][]{1975Ap&SS..35..389B, 1998SoPh..180..439W, 2008SoPh..253..173B, 
2010AdSpR..45...10B}. 
With this pressure function, the Grad-Shafranov-equation has the form
\begin{eqnarray}
\Delta A=\exp(-2A)\, , \label{lio}
\end{eqnarray}
also known as Liouville's equation \citepads[e.g.,][]{1975ArRMA..58..219B}.

By defining $u=x+iy$, $\varv=x-iy$, and $i^2=-1$, Liouville's equation can be written as
\citepads[see][]{1975ArRMA..58..219B, 1978SoPh...57...81B}
\begin{eqnarray}
4\frac{\partial^{2} A}{\partial u\, \partial \varv}=
\exp(-2A)\, .
\label{lio1}
\end{eqnarray}

The general solution of Liouville's equation, Eq.\,(\ref{lio1}), is given by
\begin{eqnarray}
A(u,\varv)=\ln\frac{1+\frac{1}{4}|\Psi(u)|^2}{\displaystyle\left|\frac{d\,\Psi}{du}\right|}\, ,
\end{eqnarray}
implying that every holomorphic function $\Psi(u)$ gives us an exact solution of
the nonlinear Grad-Shafranov-equation, Eq.\,(\ref{lio}).

The classical ansatz is
\begin{equation}
\Psi=2\exp{u}\, ,
\end{equation}
leading to the Harris-sheet equilibrium \citepads{1962IlNC...23..115H}
\begin{equation}
A=\ln\cosh (x)\, ,
\end{equation}
which represents a bipolar magnetic field structure, i.e., a plasma or current sheet
separating magnetic fields with opposite orientations. This is a one-dimensional
structure.

 \begin{figure}
   \centering
    \includegraphics[width=0.85\hsize]{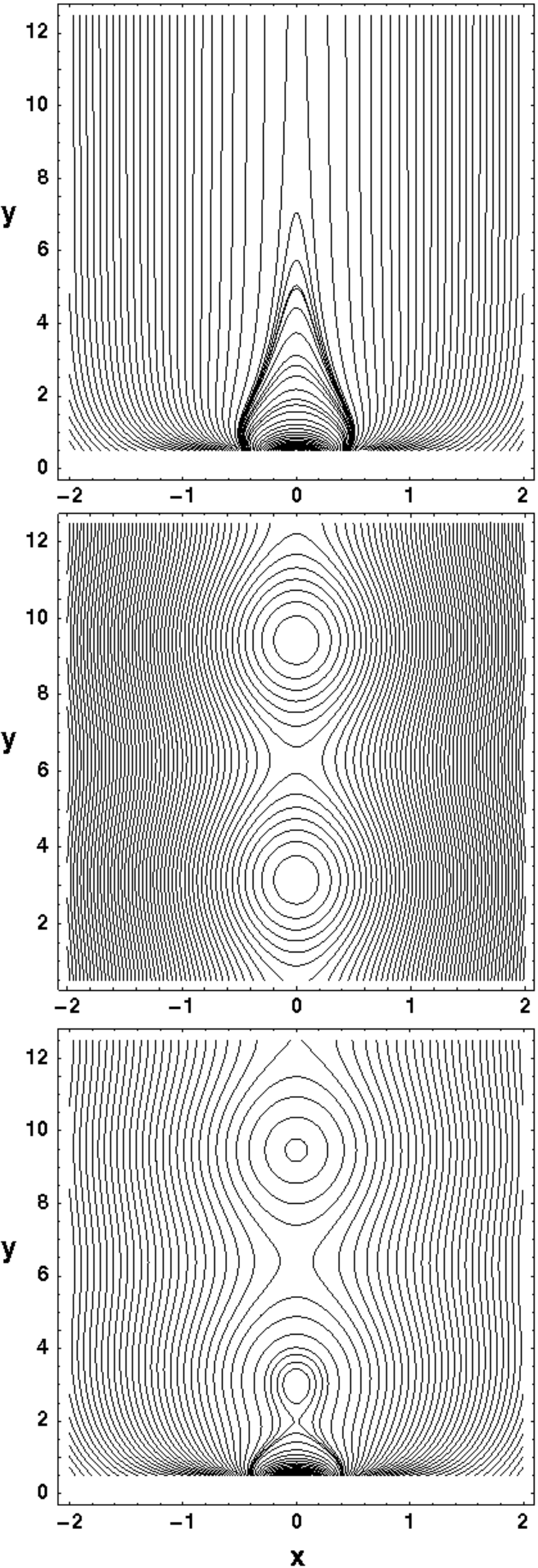}
      \caption{Field lines for the Kan magnetotail (top), the periodic sheet pinch or
       periodic Harris sheet (middle), and the combined one (bottom) that serves as our
       initial MHS equilibrium.}
         \label{fieldlinesmhs}
   \end{figure}

A modified magnetic-field structure including a normal component penetrating
the 2D current sheet was calculated by \citetads{1973JGR....78.3773K} for the case of the
Earth's magnetotail region. Such a configuration emulates the dipole structure that extends 
into and influences the Earth's magnetotail. In addition, with respect to configurations 
within the solar corona, such a scenario ideally resembles, e.g., magnetic dome structures. 
\citetads{1973JGR....78.3773K} chose the function
$\Psi=2\exp{\left(u+d/u\right)}$, which drops off for $|u|\rightarrow \infty$.
Here, $d$ is a constant. This choice is a slight perturbation of the original 1D Harris sheet 
toward a 2D magnetic-field configuration, converging for large distances back again to the 
original Harris-sheet equilibrium. The corresponding flux function is
\begin{equation}
A= \ln \displaystyle\frac{\cosh\left[ x\left(1+\frac{d}{r^{2}}\right)\right]}
{\sqrt{\displaystyle\frac{(d^{2}-2d(x^{2}-y^{2})+r^{4})}{r^{4}}}}\, ,
\end{equation}
where $r=\sqrt{x^{2}+y^{2}}$. The resulting field lines, computed for $d=0.5$,
are shown in the top panel of Fig.\,\ref{fieldlinesmhs}, with the $y$-axis pointing in
magnetotail direction, and the Earth (in the original Kan picture) located at the origin of
the coordinate system. This scenario allows one to describe stretched, tail-like structures,
including a dipole-like configuration close to the Earth (i.e., for low values of $y$).

To describe the field lines of periodic structures, we applied the approach of
\citetads{1967ApJ...149..727S}, who developed a formalism to solve Liouville's equation
resulting in the so-called periodic, corrugated sheet-pinch. In this scenario,
the original Harris-sheet equilibrium is slightly modified to $\Psi=2\left( \sqrt{1+\delta^2}
\exp{u} + \delta\right)$ with $\delta$ as a constant, leading to the following flux function:
\begin{equation}
A= \ln \left(\sqrt{1+\delta^2} \cosh{x} + \delta \cos{y}\right)\, .
\end{equation}
The field lines evaluated for $\delta = 0.1$ are shown in the middle panel of
Fig.\,\ref{fieldlinesmhs}.

For our purposes, we used the approaches from both Kan and Schmid-Burgk,
and combined them, i.e., we applied the modification of the Harris-sheet found by 
\citetads{1967ApJ...149..727S} to the Kan equilibrium.
This is necessary, because we aim at achieving a representation in which the equilibrium has a
strong $B_{x}$--component close to the lower boundary $y=0$ and a periodic-sheet pinch
for $y\rightarrow\infty$. This means that $\Psi$ is now represented by
\begin{equation}
 \Psi= 2\left(\sqrt{1+\delta^2}\exp{\left(u+d/u\right)}+\delta\right)\, .
\end{equation}
With this function, we finally obtain a flux function of the form
\begin{equation}
A = \ln\displaystyle\frac{\sqrt{1+\delta^{2}} \cosh\left[ x\left(1+\frac{d}{r^{2}}\right)\right] +
\delta \cos\left[y\left(1-\frac{d}{r^{2}}\right)\right] }
{\sqrt{\displaystyle\frac{(d^{2}-2d(x^{2}-y^{2})+r^{4})}{r^{4}}}} \, ,
\label{finalflux}
\end{equation}
whose contour plot has field lines as depicted in the bottom panel of
Fig.\,\ref{fieldlinesmhs}. In our representation the $y$-axis corresponds to the height above
the solar photosphere, and the photosphere itself is located at $y=0$. The symmetry axis of
the post-flare magnetic field configuration is given by $x=0$. This flux function
(Eq.\,\ref{finalflux}) is used in the following and serves as our initial MHS equilibrium.

\subsection{Different transformation approaches}

There exist three different approaches to model field aligned shear flows. These are the
transformations via
\begin{itemize}
\item magnetic field amplification defined by $\alpha'$,
\item peaked plasma flows defined by $M_{A}$, and
\item asymptotical 1D current structures defined by $j$.
\end{itemize}
Each approach
requires the specification of either one of the finally transformed MHD values (such as
the current or the magnetic field, the latter is even identical to the transformation itself),
or the plasma flow of the stationary MHD configuration. In combination with the prescribed
intrinsic MHS values, the corresponding transformation between these two states can be evaluated.

As we have shown, Eq.\,(\ref{Ma_alpha}) describes two equivalent methods for a transformation, 
i.e., via the calculation of either the transformed magnetic field, specified by $\alpha'$, or the
plasma flow in the transformed configuration, given by $M_A$. On the other hand, it is also
reasonable to compute the Mach number profile via Eq.\,(\ref{currenttrafoexpl6}) by specifying
the asymptotic behavior of the transformed current $j$. These three methods are basically
equivalent, but each of them emphasizes a different physical aspect of the fragmentation
problem, and consequently needs different constraints and boundary conditions.
More specifically, each method is based on the prescription of one physical parameter that
serves as control parameter.

\subsubsection{Transformation via magnetic-field amplification defined by $\alpha'$}

The first mapping method has been described in detail by \citetads{2006A&A...454..797N}.
It is based on the prescription of the magnetic field and is best applicable for potential
fields that are asymptotical homogeneous, i.e., their flux function is given by
$A\approx B_{S\infty} x$ for large $y$. In that case, the transformation between the new,
steady-state flux function $\alpha$ and the old, stationary flux function $A$ is given by
\begin{eqnarray}
\alpha(A) &=& C A + \sum\limits_{k}\, a_{k}\ln\cosh\left(\frac{A-A_{k}}{d_{k}}\right)\\
\alpha'(A) &=& C + \sum\limits_{k}\frac{a_{k}}{d_{k}}\tanh
\left(\frac{A-A_{k}}{d_{k}}\right)\, .
\end{eqnarray}
This transformation, which is based on the calculation of $\alpha'$, produces a series of $k$
Harris-sheets with different strengths $a_{k}/d_{k}$ and widths $d_{k}$, offset by $A_{k}$
from the MHS state. The parameters $C$, $A_{k}$, $a_{k}$, and $d_{k}$ are not completely 
free. They have to be chosen such that $\left|\alpha'(A)\right|>1$ to guarantee sub-Alfv\'enic 
flows and to satisfy the boundary conditions or constraints, provided, e.g., by observations.  
The number of Harris-type current sheets $k$ depends on the number of separatrix lines 
originating in potential X-points. This means that $k$ is determined or fixed by the number
of \lq pauses\rq, i.e., boundary layers, within the chosen domain. The location of the pauses 
is marked by $A_{k}$.
Such transformations are ideal for a proper modeling of tail configurations as they appear in 
the heliotail \citepads[or astrotails in
general, see][]{2001ohnf.conf...57N, 2006A&A...454..797N}, which require the maintenance of
strong current sheets that form the boundary layer in the vicinity of the seperatrix
(heliopause/astropause) in between the outer solar/stellar wind and the very local
interstellar medium.

\subsubsection{Transformation via peaked plasma flows defined by $M_{A}$}

For the second possible transformation method a Mach number profile $M_A(A)$ has to be
specified. To obtain a highly-structured current distribution, the Mach number profile
needs to contain strong gradients and must show strong spatial variation. This means that
$M_{A}$ cannot be given by a simple two-dimensional function, but has to be constructed
out of several pieces or branches, which need to be connected by continuous transitions,
meaning that each of these branches must be at least twice continuously differentiable at
the boundaries of the intervals so that no discontinuities in the current density profile
appear. Consequently, the function $M_{A}$ must be composed of a set of functions $m_{k}(A)$,
which exist only within some defined field-line interval and vanish outside. In addition,
the functions $m_{k}(A)$ must have a compact support to guarantee that both $m_{k}^{2}(A) < 1$
and $M_{A}^{2} < 1$, i.e., the Mach number and its constituents are bounded functions
(sub-Alfv\'{e}nic).

This goal can be achieved in different ways: (i) One may define piece-wise functions
that vanish at the boundary of some field-line interval; (ii) one may apply purely the
classical partition of unity; (iii) the partition of unity is used in combination with a
continuous sum, i.e., continuous distribution of flow tubes that can be written as an integral.
This means that we can define a general function for $M_{A}$ of the form
\begin{eqnarray}
M_{A}(A)=\sum\limits_{k} m_{k}(A) + \int\limits m(A)\, dA \,\, ,
\end{eqnarray}
which consists of a sum over $k$ individual peaked plasma flows, each defined within
discrete field-line intervals, and a continuous distribution $m(A)$ of plasma flow tubes.

Such an approach with a pure continuous distribution $m(A) \sim 1/\cosh(A)^{2}$ has been used,
e.g., by \citetads{2010AnGeo..28.1523N} to generate a single current-sheet along a magnetic
separatrix. In contrast, we show in Sect.\,\ref{example} an example in which piece-wise
functions are defined.

\subsubsection{Transformation via asymptotical 1D current structures defined by $j$\,: the inverse method}
\label{inversemethod}

The third method to determine the transformation is based on the prescription
of the transformed current, $j$, given by Eq.\,(\ref{currenttrafoexpl6}) \citepads[see
also][]{2006A&A...454..797N}. Typically, in  magnetostatics
the magnetic field is directly calculated from Amp\`{e}re's equation $\Delta A
= -j$, where the current distribution $j$ is prescribed. In MHD, a prescription of the
current  distribution or the magnetic field is not possible. Here, the values have to be
calculated self-consistently and simultaneously from the nonlinear MHD equations.
But the transformation method enables us to define the current distribution in
configurations that occur ubiquitously in space plasmas as an explicit function of the flux
function $A$. If we can find a way to prescribe the current density $j$
as a spatially, i.e., depending on $A(x,y)$, strongly variable current distribution,
we can generate self-consistently current fragmentations on small scales, emulating
scenarios that in the literature are often approached via turbulence originating
from external perturbations
\citepads{1999ApJ...517..700L, 2009ApJ...700...63K, 2011PhRvE..83e6405E}.

As was shown in Sect.\,\ref{trafo}, the current distribution, resulting from the
transformation, is given by
\begin{equation}
-j=\Delta\alpha=\alpha'' \left(\vec\nabla A\right)^2 + \alpha' \Delta A \, ,
\label{currenttrafoexpl1}
\end{equation}
with $j=j_{z}(x,y)$. The terms $\alpha'', \alpha'$, and $\Delta A$ are pure functions of the
flux function $A$. On the other hand, the quadratic expression $\left(\vec\nabla A\right)^2$
is a scalar function and generally a function of $x$ and $y$, which, in the case of 2D
equilibria, cannot be expressed as an explicit function of $A$ only. Instead,
Eq.\,(\ref{currenttrafoexpl1}) is an equation defining or rather determining
$j(x,y)\equiv j(A,y)$ from a given transformation $\alpha'$, which seems to be the most
consequent and logical method. However, Eq.\,(\ref{currenttrafoexpl1}) cannot be regarded as
just a pure ordinary differential equation for $\alpha'$. Therefore, calculating $\alpha'$
from Eq.\,(\ref{currenttrafoexpl1}) for a given or prescribed $j$, has in general no formal
solution (see Appendix\,\ref{append}). Based on this mathematical problem, it is necessary to
find a different approach.

Magnetohydrostatic equilibria in space plasmas often have regions where the fields are
extremely stretched. Such tail-like regions typically occur far away from bipolar or even
multipolar field regions, as, e.g., in our case (bottom panel of Fig.\,\ref{fieldlinesmhs})
in the regions of high $|x|$ values, or, in the case of the Kan equilibrium, also in the
regions of high $y$ values (top panel of Fig.\,\ref{fieldlinesmhs}), or in general for going
to $\infty$ along or in the direction of the tail axis. The regions of stretched field lines
can be approximated by a 1D configuration, which depends only weakly on a second coordinate.
Examples are asymptotically 1D
regions of exact and analytical tail equilibria or so-called weakly 2D or weakly 3D equilibria
\citepads[see, e.g.,][]{1972ASSL...32..200S, 1977JGR....82..147B, 1979SSRv...23..365S}. The
advantage of this asymptotically 1D approach is that the equilibrium problem can be treated
as if it depended on only one coordinate. This coordinate is, at least locally, a unique
function of the field-line label $A$, and vice versa, so that in a local coordinate system $A$
can always be chosen as coordinate. Therefore, in these stretched field line regions, the
problem can be solved. The solution found is, however, a general solution and not restricted
to the pure 1D region, because every found solution for $|\alpha'| > 1$ is an exact solution
of the sub-Alfv\'{e}nic steady-state problem (Eqs.\,\ref{magtrafo}-\ref{mhs1}).

Assuming that $\lim_{x,y\rightarrow\infty}\vec B_{S}= \vec B_{S\infty}$, we define
$|\vec B_{S\infty}|=B_{S\infty}(A)$. The limes has to be smooth. Then we can
introduce the asymptotical current via $\vec\nabla\times(\lim_{x,y\rightarrow\infty}
\vec B_{S})=\lim_{x,y\rightarrow\infty}(\vec\nabla\times\vec B_{S})=\lim_{x,y\rightarrow\infty}
\vec j_{S}$, with $\lim_{x,y\rightarrow\infty}|\vec j_{S}|=j_{S\infty}(A)= P_{S} \, '(A)$.
The last identification thereby again represents the Grad-Shafranov-equation (Eq.\,\ref{GSE}).
With these relations, Eq.\,(\ref{currenttrafoexpl1}) can be written as
\begin{eqnarray}
- j_{\infty}(A)=\alpha'' (A) B_{S\infty}^2(A) - \alpha'(A)\, j_{S\infty}(A)\, .
\label{currenttrafoexpl3}
\end{eqnarray}
This pure one-dimensional differential relation is now a linear ordinary differential
equation of first order for $\alpha'(A)$, which can be solved: We divide both sides of
Eq.\,(\ref{currenttrafoexpl3}) by $B_{S\infty}^2$ and multiply with the integrating factor
$\exp\left(\int (- j_{S\infty}/B_{S}^2)\, dA\right)$. This leads to a complete differential
that can be integrated, resulting in the following general solution
\begin{eqnarray}
\alpha'= \alpha'(A)=\displaystyle\frac{\displaystyle\int\exp{
\left(\int\,\displaystyle -\frac{j_{S\infty}}{B_{S\infty}^2} \, dA \right)}
\left(-\frac{j_{\infty}}{B_{S\infty}^2}\right) \, dA + C_{0}}
{\exp{\displaystyle\left(\int\, -\frac{j_{S\infty}}{B_{S\infty}^2} dA\right)}} \, .
\label{currenttrafoexpl4}
\end{eqnarray}
Thus for reasonable prescribed stationary asymptotic current densities $j_{\infty}(A)$, the
general solution of the transformation $\alpha'$ can be computed from the asymptotic MHS
functions for the current density $j_{S\infty}(A)$ and the magnetic field $B_{S\infty}(A)$.
The parameter $C_{0}$ is an integration constant, defining an offset and a boundary condition
for $\alpha'$, hence $M_{A}$.

To find suitable prescriptions for $j_{\infty}(A)$ that guarantee $\alpha'^{2}>1$ also in the
two-dimensional regions is a difficult practical task. However, with this approach it is
basically possible to generate current fragmentation scenarios based purely on the
self-consistent solution of the MHD equations. Hence, the transformation method provides a
self-consistent tool in which current fragmentation is a basic property of the nonlinear
MHD theory, because $j_{\infty}$ is basically not subject to any limitations.

\subsection{Example for shear-flow-induced current fragmentation}
\label{example}

\begin{figure}
  \centering
   \includegraphics[width=\hsize]{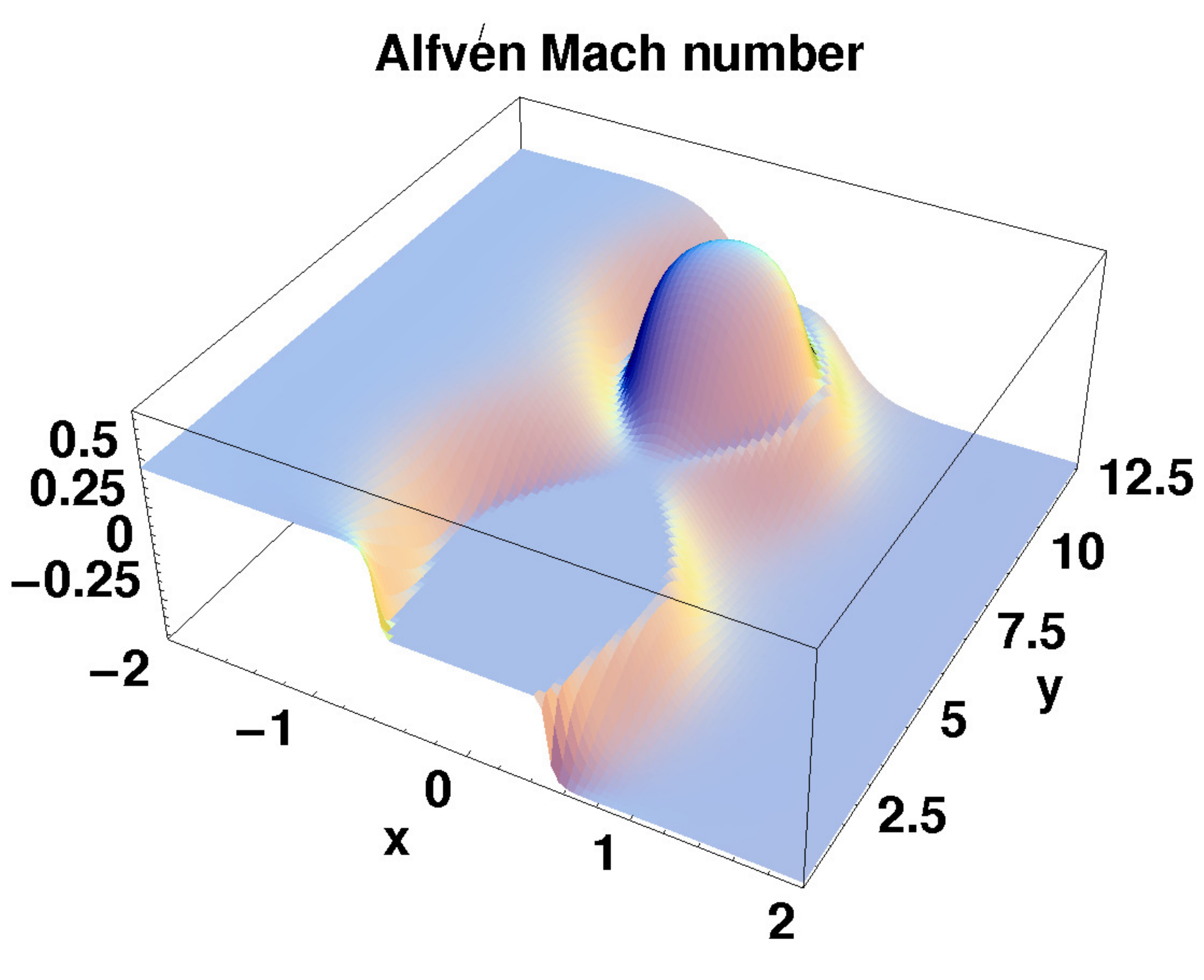}
      \caption{Constructed Mach number profile.}
         \label{machnumber1}
   \end{figure}

The aim of our analysis is to study the process of current fragmentation that takes place in the
vicinity and within an ejected plasmoid that was formed via magnetic reconnection in a
typical solar eruptive flare. Observations of such flare processes indicate that the
surrounding material on the open field-lines is moving upwards, while the plasma below the
X-point located in between the two plasmoids (see Fig.\,\ref{fieldlinesmhs}), i.e., within
the closed field-line region, can be assumed to be static\footnote{In dynamical flare 
scenarios, the plasma within the closed arcade structure 
tends to flow downwards, i.e. back to the surface because of plasma cooling, which forces the
arcade structure to shrink. However, our model aims at studying the post-eruptive flare 
phase, in which the photospheric layers (i.e., the arcade structures at the bottom of our 
configuration) are back in static equilibrium \citepads[see, e.g.,][]{2012ApJ...757L...5W}.}.
With this picture, it is more
convenient to apply the transformation method based on the prescribed nonzero
sub-Alfv\'{e}nic Mach number profile rather than based on the asymptotic current
distribution or the magnetic field amplification, because the latter two are extremely
difficult to extract from observations, in particular because of the still-lacking high
enough spatial resolution.

In a bipolar MHS structure, assuming the main direction of the magnetic field to be the
$y$--direction, i.e., $B_{y}>B_{x}$ outside the outer separatrix, the main component of the
magnetic field, $B_{y}$, changes its direction and therefore its sign. For
symmetric magnetic-field lines with respect to the $y$-axis, $A$ is a symmetric function of
$x$ (i.e., $\vec B_{S}$ is anti-symmetric and $A$ is symmetric with respect to the $y$--axis).
As $M_{A}$ is a function of $A$, and the plasma flow is required to be purely upstreaming
on both sides (boundary condition), the Mach number profile needs to change its sign.
Consequently, one needs to define a piecewise function $M_{A}(A)$ with at least two
different branches (left and right of the outer separatrix). Otherwise, $M_{A}$ cannot change
its sign.

\begin{figure*}
  \centering
   \includegraphics[width=\hsize]{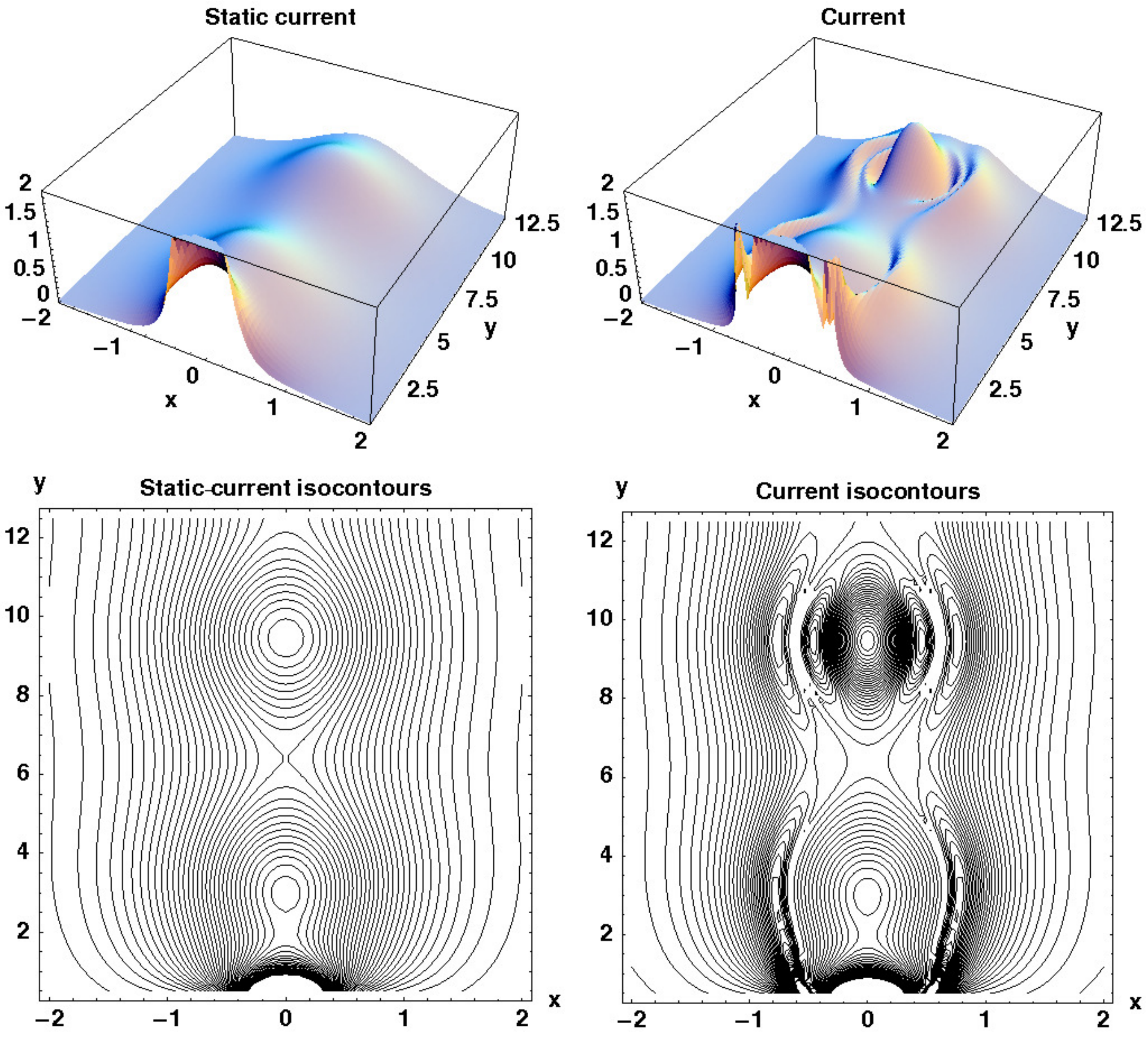}
      \caption{Static (left) versus stationary (right) current (top)
and its isocontours (bottom). For better visualization the current is plotted inversely and
cut off at the numerical value of 2. The maximum at the origin approaches a numerical 
value of 6.}
         \label{mappingfigs}
   \end{figure*}

The Mach number profile, which we apply now to simulate such a scenario, consists of several
branches: Within the dipole-like region inside the closed field-lines, which is assumed to be
static, $M_{A} = 0$. In the regions outside the outer separatrix, we assume that the Mach
number profile is symmetric, i.e., $M_{A}(-x,y)=-M_{A}(x,y)$, although asymmetric profiles
might also be possible. We furthermore assume that for $x<0$ the plasma flow is parallel to
the magnetic field ($M_{A} > 0$), while for $x>0$ it is antiparallel ($M_{A} < 0$). To
guarantee a continuous transition between the positive and negative Mach number branches
(i.e., to have a smooth flow pattern), $M_{A}$ must vanish at the separatrix itself. To
ensure that the current $j$ is also continuous, $M_{A}$ and its derivative must be
continuously differentiable in the boundary region (separatrix). Furthermore, we assume that
the upper, disconnected plasmoid contains a vortex. This assumption is reasonable,
because any small asymmetry during the reconnection process and the disconnection of the
plasmoid will immediately result in a nonzero angular momentum and hence in a rotational
motion of the plasma. Hence, its representation in the Mach number profile is given by a
maximum in the center of the plasmoid, and strong gradients from the center to its edges.

With these specifications, our Mach number profile covering the region in $x$ and $y$ as
defined by the MHS configuration (see Fig.\,\ref{fieldlinesmhs}) is given by the following
four branches
\begin{displaymath}
M_{A}  =
\left\lbrace \begin{array}{lcl}
-0.5 f \left( 1-\displaystyle\frac{1}{1+ \left(\frac{A-A_{\rm sep}}{A_{b}}\right)^{2} }\right) & \textrm{for} &
 A > A_{\rm sep}\, ,\quad x > 0 \\
0.5 f \left( 1-\displaystyle\frac{1}{1+ \left(\frac{A-A_{\rm sep}}{A_{b}}\right)^{2} }\right) & \textrm{for} &
 A > A_{\rm sep}\, ,\quad x < 0 \\
 0.9 f_{p}\left( 1-\displaystyle\frac{1}{1+ \left(\frac{A-A_{\rm sep}}{A_{b}}\right)^{2} }\right) & \textrm{for} &
 A < A_{\rm sep}\, ,\quad y > 6.5 \\
 0 & & \textrm{elsewhere}\, ,
\end{array}
\right.
\end{displaymath}
it is displayed in Fig.\,\ref{machnumber1}. Hereby $A_{\rm sep}$ represents the outer
separatrix and has the numerical value $A_{\rm sep}=0.0875$,
and $A_{b}$ is a parameter influencing the steepness of the Mach number profile, and therefore
the width of the current sheets. For our model computations we choose $A_{b}=0.1$. The
parameters $f$ and $f_{p}$ are functions of $A$, simulating small wave-like spatial fluctuations.
For the example presented in Fig.\,\ref{mappingfigs}, we used $f = 1- 0.1 \sin{(1.1 A)}$ and
$f_{p}=1$.

Starting from the MHS equilibrium configuration for the flux function and its corresponding
current distribution (see Sect.\,\ref{nonlinstat}), we applied the mapping defined by the
Mach number profile. The resulting current and its isocontour lines are displayed in
the upper and lower right panels of Fig.\,\ref{mappingfigs}. Obviously, the current
distribution shows new features, which did not exist in the static case (left panels of
Fig.\,\ref{mappingfigs}). These are ring-like and crescent-shaped structures around both the
lower, static configuration and the upper disconnected
plasmoid. In both cases, these new current sheets are located outside but along the
separatrix. In addition, inside the detached plasmoid, the current appears dome-like in the
center, and two more current sheets (current \lq islands\rq) formed close to the separatrix.
These new current structures (sheets, islands, maximum) are
also visible in the isocontour plot. There, additional butterfly-like current islands
appear in the vicinity of the X-point of the outer separatrix.
The strongest currents can be recognized at the \lq Kan dipole\rq-region,
because the increase of $A$ and $|\vec B_{S}|$ in the region of the pole of $A$ results in
strong currents already in the MHS-state. Applying the shear at the outer separatrix
enforces this effect in particular at the bottom separatrix (in the vicinity of the
photosphere), generating two additional current peaks.
In the center of the static configuration (where $M_{A}$ was set to zero),
the current structure has not changed. In contrast to the Grad-Shafranov theory, the stationary
current isocontours do not (completely) resemble the field-line structure anymore.

\begin{figure}
     \includegraphics[width=\hsize]{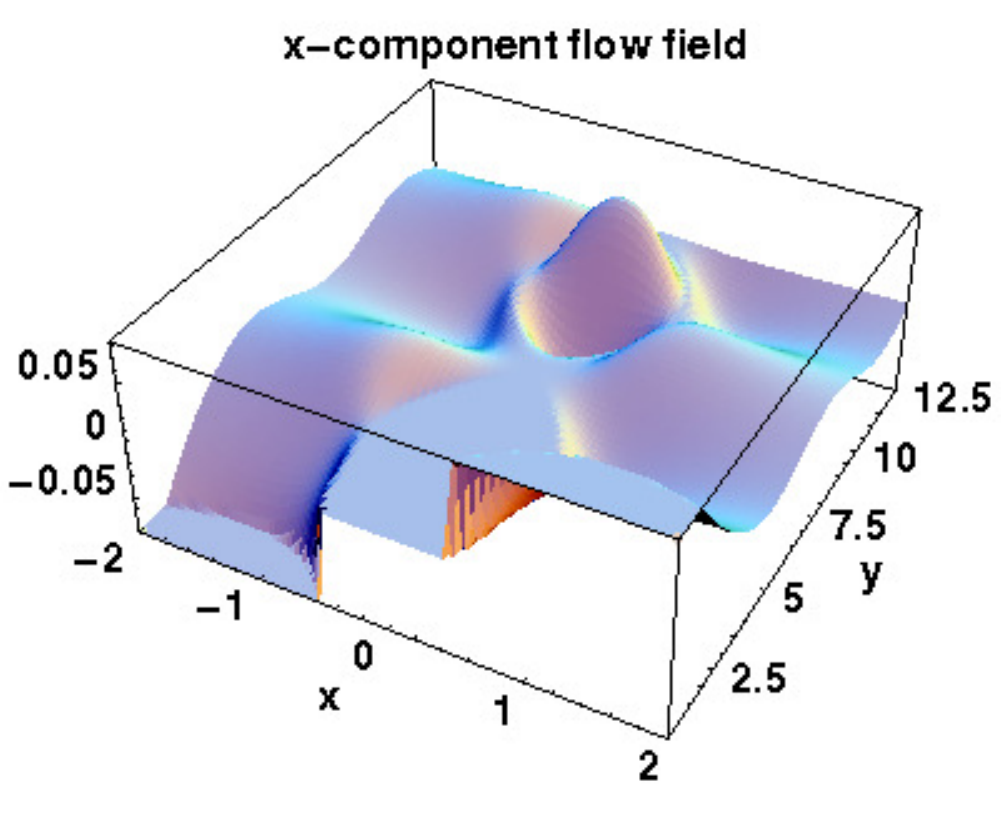}
     \includegraphics[width=\hsize]{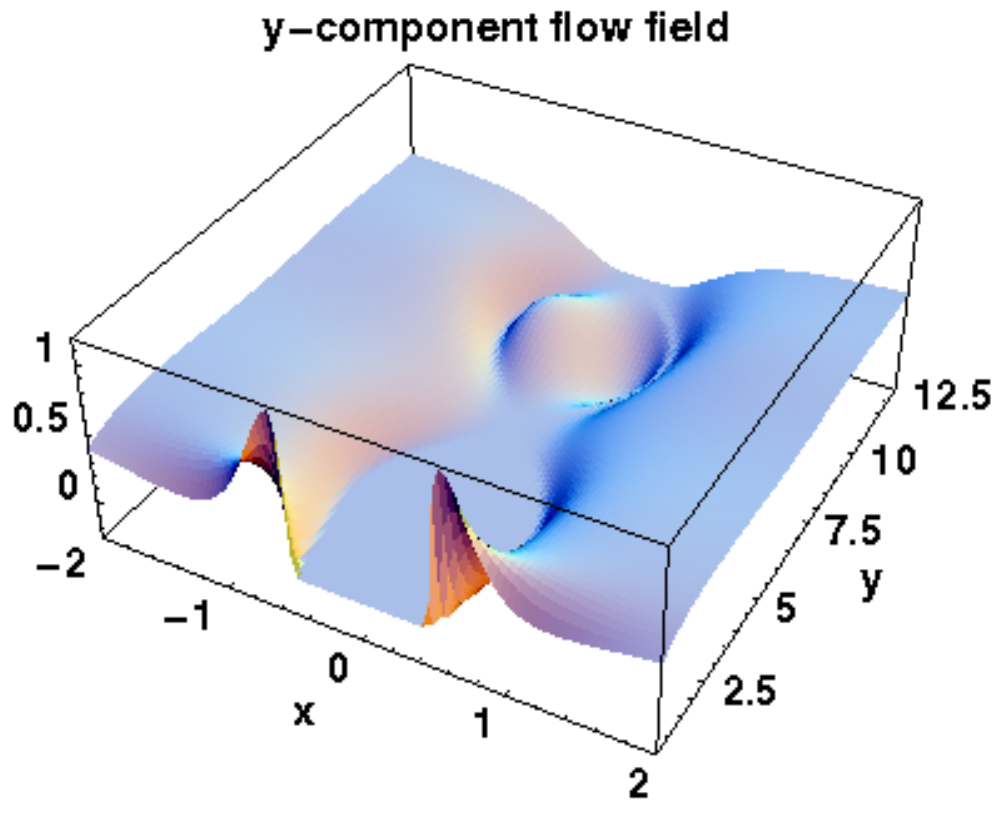}
 \caption{$x$ and $y$ components of the plasma flow field.}
         \label{flowxy}
     \end{figure}

To highlight the motion of the plasma, we display in Fig.\,\ref{flowxy} the $x$ and $y$
components of the normalized (with respect to density) plasma velocity field. The $y$
component shows that in the outside regions it is always positive, in agreement with an
upstream behavior of the flow on the open field-lines. The clockwise rotational flow of the
plasma within the upper plasmoid is obvious from the $y$ component of the flow being positive
on the left side of the plasmoid, and negative on the right side. The $x$ component of the
flow is generally very small with a wavy structure due to the curved open field-lines in the
vicinity of the outer separatrix. Only the circular flow inside the upper plasmoid has
slightly higher velocity.

In summary, from an initially smooth current
distribution our applied mapping created a new distribution, which shows multiple current
filaments that can be regarded as current fragmentation.

 \begin{figure}
     \includegraphics[width=\hsize]{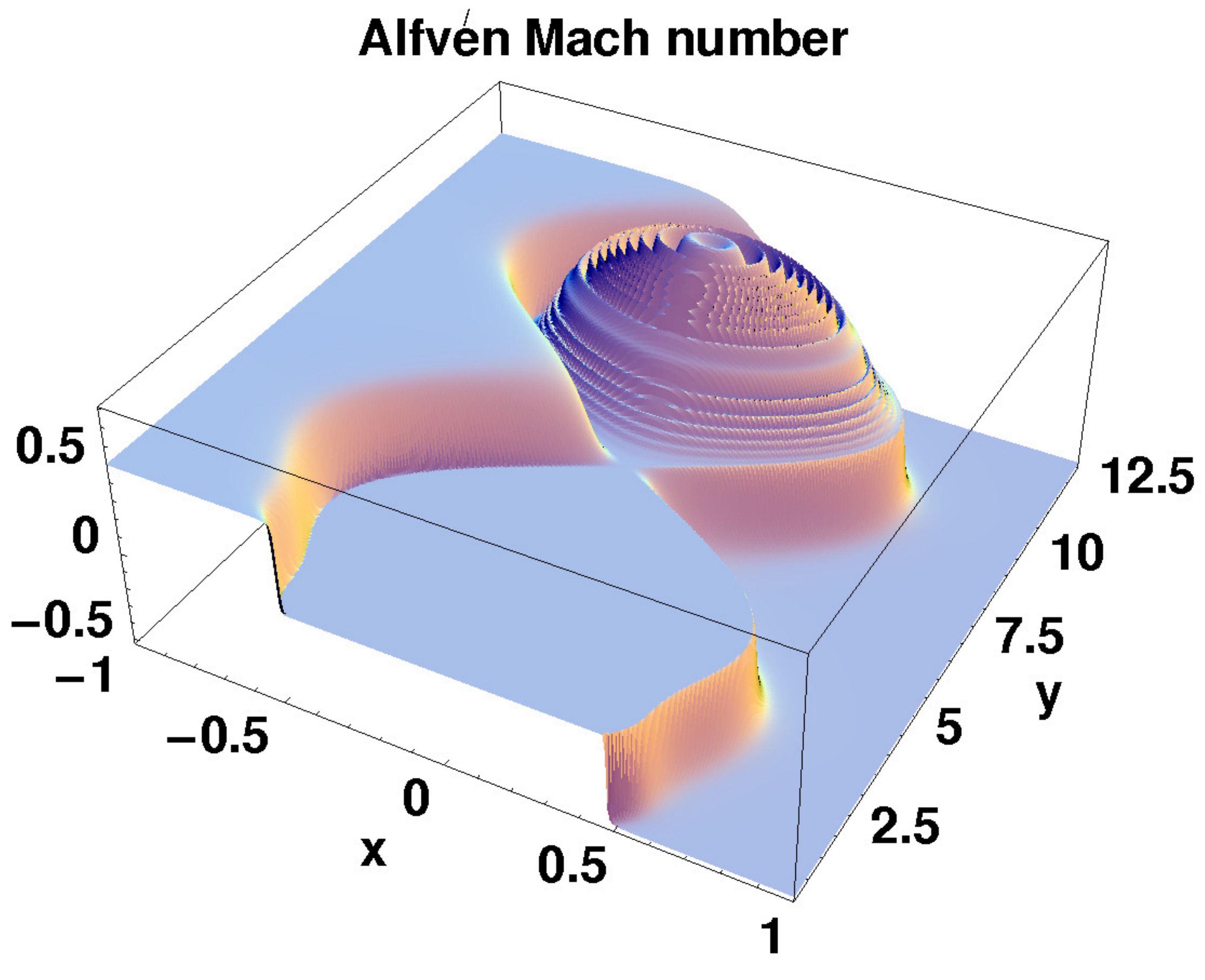}
     \includegraphics[width=\hsize]{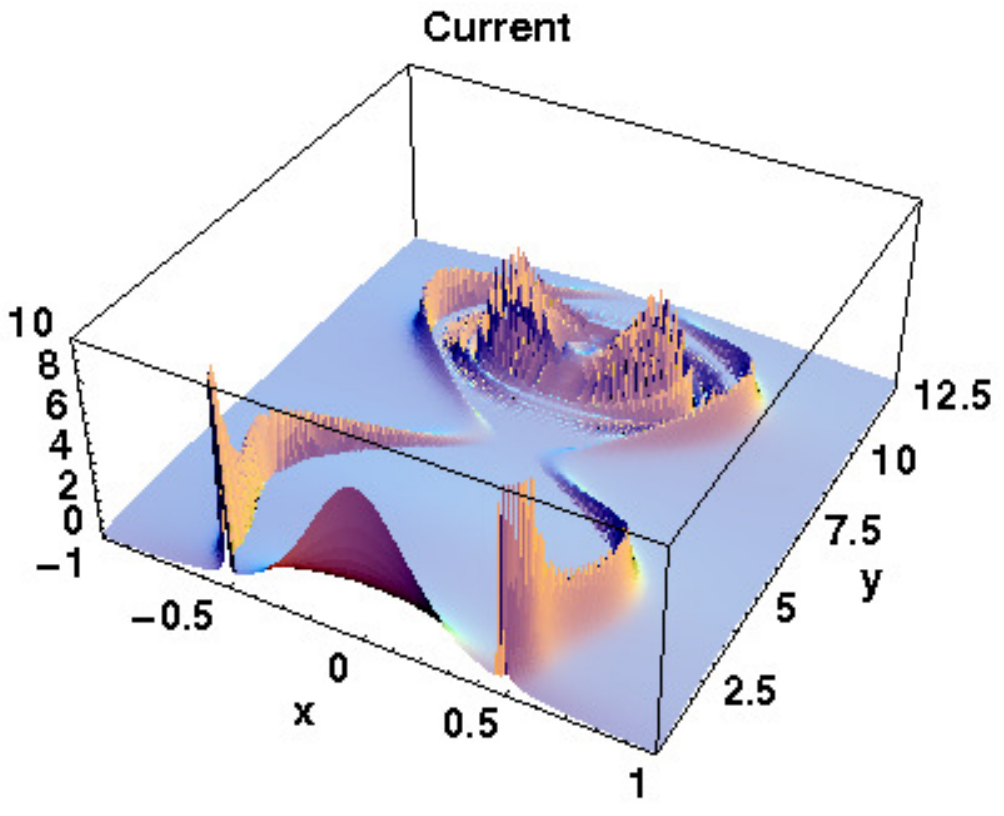}
     \includegraphics[width=\hsize]{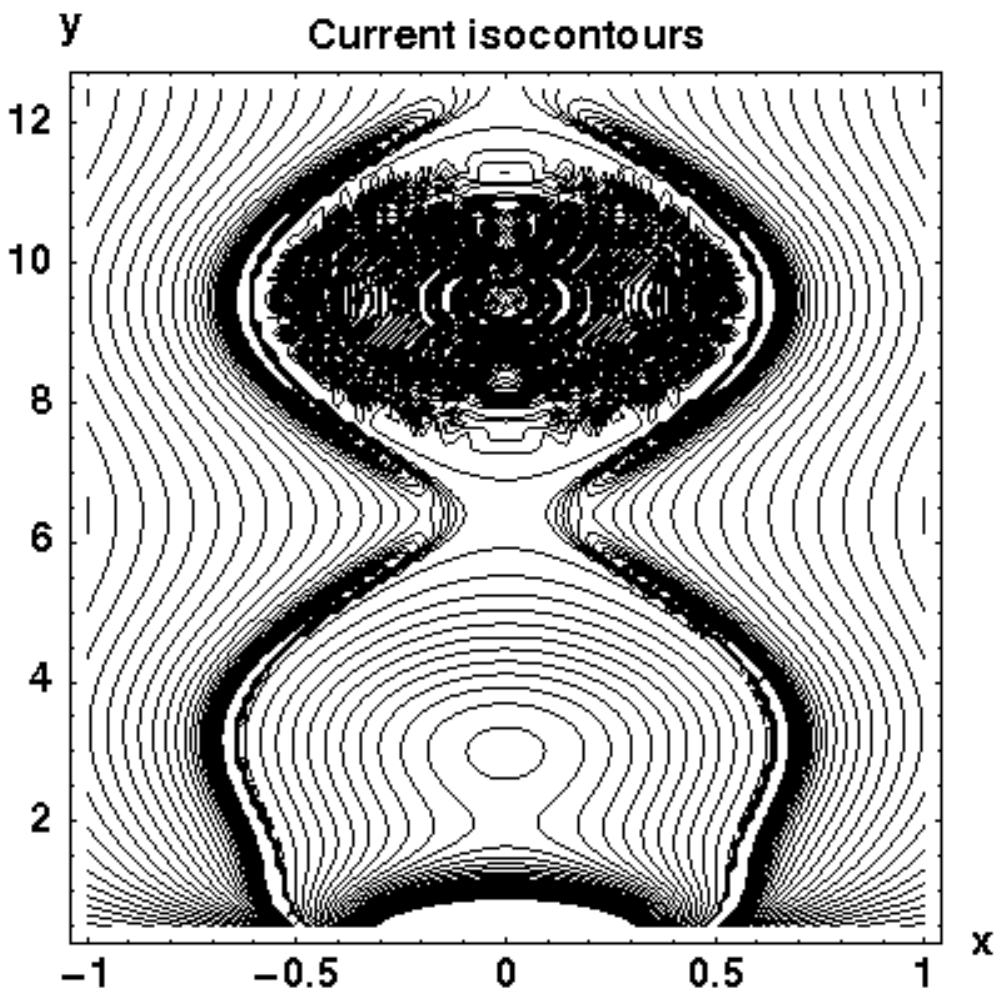}
 \caption{Mach number profile (top), transformed inverse current density (middle) and its
         isocontours (bottom), for a narrower intrinsic current sheet with  $A_{b} = 0.01$ 
         and an intrinsically structured plasmoid with $f_{p} \neq 1$.}
         \label{Ab001}
     \end{figure}

\section{Discussion}
\label{sec4}

\begin{figure}
     \includegraphics[width=\hsize]{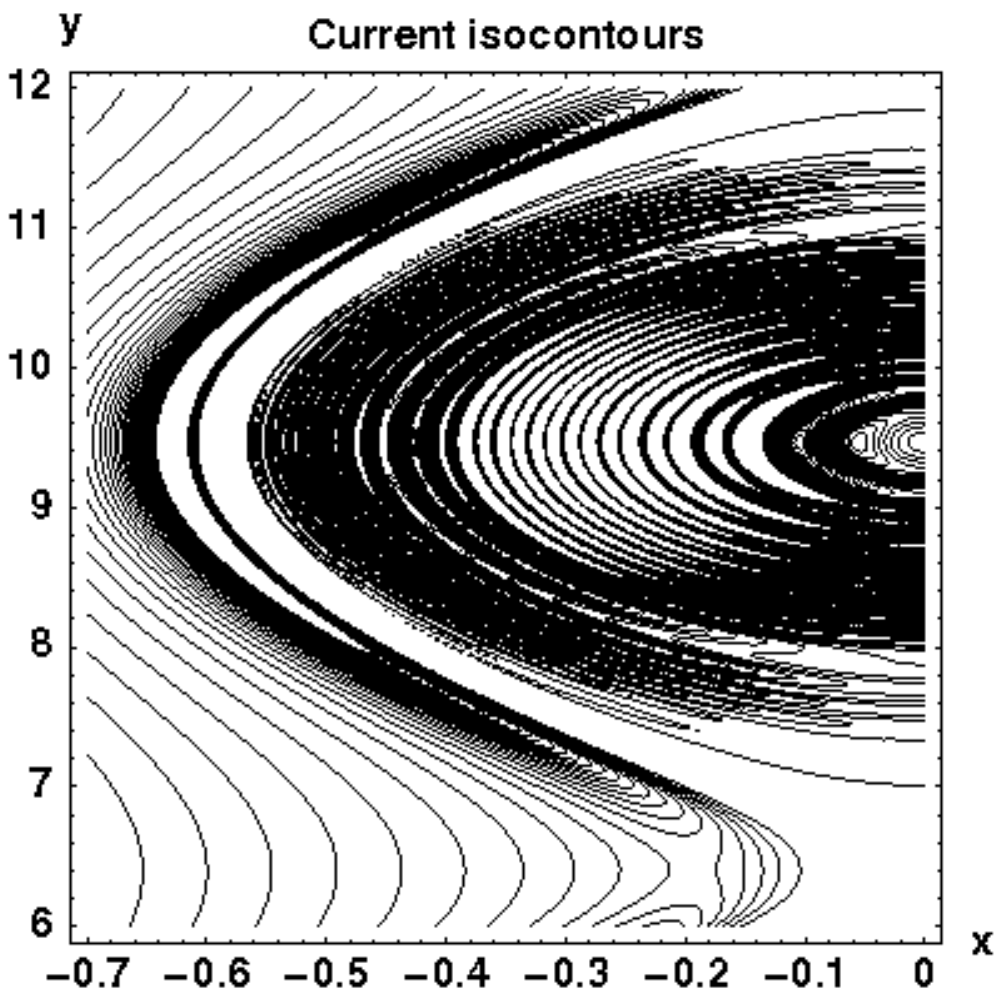}
     \includegraphics[width=\hsize]{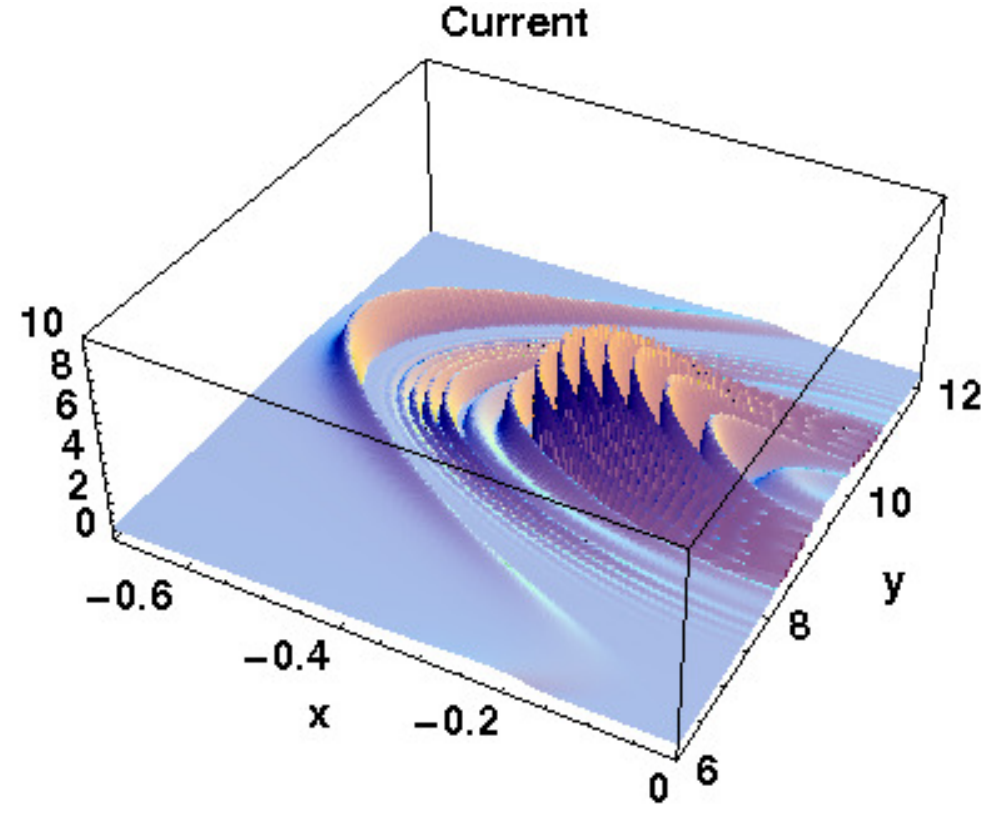}
     \includegraphics[width=\hsize]{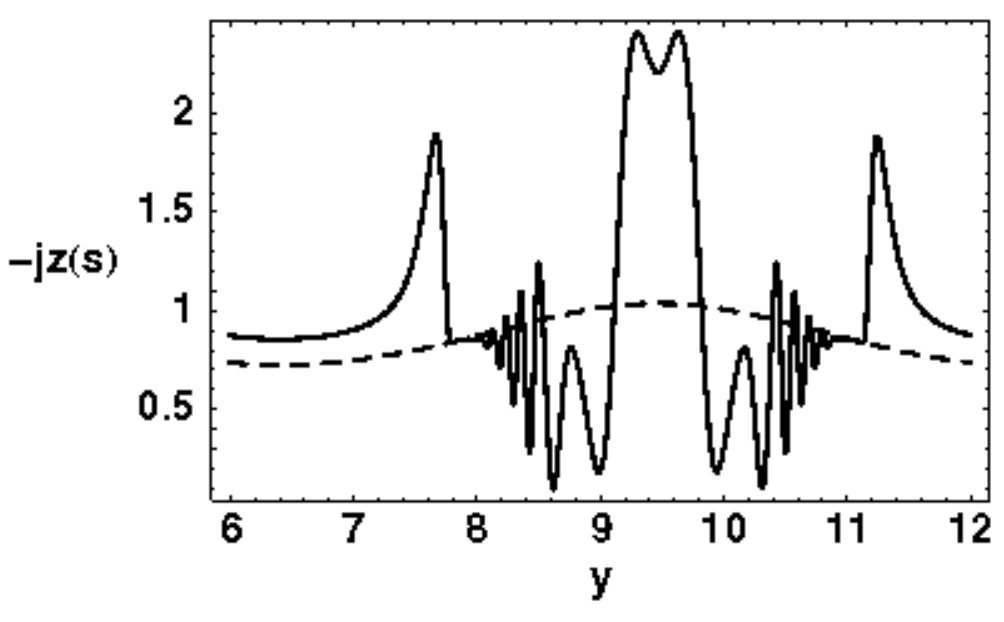}
 \caption{High-resolution zoom into the plasmoid region for the same model as in 
   Fig.\,\ref{Ab001}. Shown are the isocontours (top), the transformed inverse current 
density (middle), and a cut through the current density at $x=-0.4$ (bottom) for the 
stationary ($j_{z}$, solid) and the static case ($j_{zs}$, dashed).}
         \label{Zoom}
     \end{figure}

The results shown in the previous section serve as an illustration. There, the width
of the current sheet, prescribed by the parameter $A_{b}$, was set to a value of 0.1, which
resulted in pure, crescent-shaped current sheet structures. However, decreasing the value of
$A_{b}$, i.e., steepening the Mach number profile and shrinking the width of the current sheet 
resulted in additional fragmentation of the crescent-shaped current sheet into several strong current 
peaks, as shown in the middle panel of Fig.\,\ref{Ab001} for which $A_{b}$ was set to 0.01. For 
better visualization the current was cut at the numerical value of ten. From these results we may 
thus conclude that fragmentation is enforced when reaching smaller scales of the shear flows.
Furthermore, fragmentation of the flux rope's current density profile occurs when the parameter
$f_{p}$ is different from 1. In the example depicted in Fig.\,\ref{Ab001} we used
$f_{p} = 1-0.1\sin[1/(A^{2}+0.01)]$ and $f=1$. The corresponding Mach number profile, which now
already shows a small-scale structure imprinted on the plasmoid, is shown in the top panel of
Fig.\,\ref{Ab001}. The fine-structure obtained in the transformed current is obvious from
both the current density profile and its isocontours (bottom panel of Fig.\,\ref{Ab001}). 
For better visualization we also show in Fig.\,\ref{Zoom} a high-resolution zoom of the 
isocontours and the inverse current density for the same model parameters as in 
Fig.\,\ref{Ab001}. The zoomed region contains the left side of the plasmoid. The plot of
the isocontours demonstrates that the topology of the current isolines is much more 
complex than the one of the flux function isolines, i.e., the magnetic-field lines. 
The bottom panel of Fig.\,\ref{Zoom} displays the comparison between the static 
(dashed lines) and the stationary, fragmented current density (solid line) along the $y$-axis
within the plasmoid region for $x=-0.4$. This plot highlights the strong spatial variation of 
the stationary current density, which shows steep gradients that imply fragmentation of the 
initially smooth current sheet. This relatively simple example stresses that to achieve 
fragmentation on much smaller scales, it is essential to use a Mach number profile, which is much 
more complex and contains highly alternating structures, e.g., in the form of saw-tooth-like or 
other oscillating functions. Furthermore, every Mach number profile could in principle successively 
and infinitely be refined by, e.g., an iterative scheme of the form $f((M_{A})_{n}) = 
(M_{A})_{n+1}$. Such an iterative mapping can be performed because $(M_{A})_{n}$ is constant along 
the field lines, so every regular function or mapping has to be constant on the field lines as 
well. These iterations define fractal structures and hence demonstrate the fractal nature of MHD.
The concept of fragmentation in the frame of ideal MHD remains valid down to length scales of 
100\,m to $\sim 5$\,m for conditions typical for the solar corona. The value of $\sim 5$\,m 
thereby corresponds to the ion inertial length, defined as the ratio of the speed of light
and the ion plasma frequency. On length scales similar to and shorther than the ion 
inertial length, either the Hall-MHD or the two-fluid MHD needs to be applied.

In our analysis we ignored resistive or nonideal effects to guarantee the existence
of plausible stationary flows. However, the presence of nonideal terms, particularly
in the shape of a resistivity on the right-hand side of Ohm's law, does not automatically
imply the nonexistence of stationary solutions. The inclusion of a resistivity, $\eta$,
such that $\vec\nabla\times \left( \eta \vec j\right) = \vec 0$, supports stationary
nonideal MHD flows and hence the existence of ideal equilibria. The stationarity
of Maxwell equations in 2D demands that the electric-field component $E_{z} = \eta j_{z}$ is
constant. As $E_{z}$ is at the same time the reconnection rate, this implies that the
reconnection rate is independent of the resistivity \citepads[e.g.,][]{2006PhPl...13c2307K}.
Consequently, even if, as in our case, the flows are field-aligned and steady-state, these
MHD flows can be regarded as an analogy to steady-state reconnection solutions with constant
reconnection rate. The existence of resistive steady states with field-aligned flows and
reasonable resistivity profiles has been shown by \citetads{2000JPlPh..64..601T,
2003PhPl...10.2382T}. Under such conditions in our 2D case, Ohmic heating of the plasma is
directly proportional to $j_{z}$ and occurs everywhere where filamentation or fragmentation
takes place and could in principle contribute (at least partially) to the heating of the
corona.

Although we had limited our analysis to a pure 2D configuration, the transformation technique
is valid in all dimensions because it is based on vector analysis identities. Therefore, starting 
from a 3D MHS equilibrium, the mapping would deliver current fragmentation also in 3D. However,
to find suitable MHS equilibria as starting configurations is a difficult task, hence 
pre-computed fully 3D MHS equilibria are so far rare. MHS equilibria for laminar 
flows and magnetic fields have been constructed by, e.g., \citetads{1999GApFD..91..269P}, which 
might serve as starting points for a future 3D analysis.

\section{Conclusions}
\label{sec5}

Observations of the solar atmosphere with increasing spatial resolution reveal that the atmosphere
is highly structured or fragmented. Hence, the mechanisms initiating the formation of small-scale
structures, such as jets, flares, and plasmoids, which typically occur as a result of magnetic
reconnection processes of current sheets, must be inherently fractal.

Although solutions of the MHD equations pretend that physical parameters, such as the magnetic 
field or the current, are smooth on large scales, they do not necessarily have to be smooth on small
scales. This is shown by our analysis, in which, starting from an MHS equilibrium with a smooth
current distribution for a stationary plasmoid configuration, we obtained a current structure
displaying steep gradients, i.e., strong spatial variations of the current density, as well as
an internally fragmented plasmoid, depending on the initially chosen Mach number profile.
Hence, pure MHD equilibria are able to display intrinsic fine structure, which
can serve as the seeds for instabilities, i.e., as ``secondary instabilities''
\citepads[see, e.g.,][]{2010AIPC.1242...89P}, and therefore as triggering mechanisms for
second-generation current fragmentation.

Because the MHD equations are scale-free, our results are valid
not only for the global flare scale, but also for scales close to dissipation
scales.

As a natural next step, our stationary equilibrium configuration should be implemented into
MHD simulations as the starting configuration, to see and test the onset of instabilities and the
time-dependent evolution of the resulting additional current fragmentation.

\appendix

\section{Proof of theorem}    
\label{append}

In Sect.\,\ref{inversemethod} we claimed that for a given $j$
Eq.\,(\ref{currenttrafoexpl1})
has in general no formal solution. One may argue that it is always possible to reduce
Eq.\,(\ref{currenttrafoexpl1}) to an ordinary differential equation for $\alpha$ as a
function of $A$. Here we show that this is indeed not the case, because the solution to any such
differential equation returns the original form of the equation.

An equivalent representation of the current transformation equation
(\ref{currenttrafoexpl1}) would be
to use instead of the coordinates $x$ and $y$ the flux function $A$ and the arc length $s$
along a field line, or instead of $s$ one of the coordinates $x$ and $y$. The choice of
such a representation has pure mathematical reasons: Eq.\,(\ref{currenttrafoexpl1})
should present an ordinary differential equation for $\alpha$, and $\alpha$ itself
should depend only on one single coordinate $A$. But the nonconstant coefficients of
$\alpha'$ are depending on two coordinates. The choice of a coordinate system that includes
$A$ as one of the coordinates enables us to formulate a constraint for which current
distributions $j$ the Eq.\,(\ref{currenttrafoexpl1}) is really an ordinary differential equation
for $\alpha$ as a function of $A$.

As long as $A$ is locally monotonic, $B_{S}^2$ usually depends explicitely on the flux
function $A$ and the arc length $s$ along a field line as coordinates equivalent to $x,y$ in
2D, i.e. $B_{S}^2\equiv \left(\vec\nabla A\right)^2(x,y)\equiv \left(\vec\nabla
A\right)^2(A,s)\equiv \left(\vec\nabla A\right)^2(A,y)\equiv$ etc.

Taking the function $j(A,s)$ as a constraint for the MHD system,
one has to solve a partial differential equation in $A,s$ or $x,y$. The special shape
\begin{equation}
-j(A,s)=\alpha''(A) B_{S}^2(A,s) - \alpha'(A)\, P_{S}\, '(A)
\label{currenttrafoexpl2}
\end{equation}
is a strong restriction for every prescribed current (function) $j(A,s)$.
How to solve it correctly for any arbitrarily prescribed $j(A,s)$? The problem is caused by
the fact that Eq.\,(\ref{currenttrafoexpl2}) is an equation {\it defining} or
rather {\it determining} $j(A,s)$ from a {\it given} transformation. Therefore, to prescribe
$\alpha'(A)$ to calculate $j(A,s)$ seems to be the most consequent and logical method.

Nevertheless, an inverse method for calculating the transformation $\alpha(A)$
or rather $\alpha'(A)$ from the transformation equation for the current
(Eq.\,(\ref{currenttrafoexpl2})) would have great advantages, because it would make it
possible to generate current fragmentation and strong current-sheets of arbitrary,
\lq turbulent\rq~structure, needed to induce magnetic reconnection or general plasma
instabilities.

However, there are several obstacles to use an inverse method. First, it will be
difficult to express $B_{S}^2(x,y)$ as $B_{S}^2(A,s)$, at least analytically.
Second, one has to assume a special dependence of the current density
$j(A,s)$ to fulfill Eq.\,(\ref{currenttrafoexpl2}). But any \lq arbitrary\rq~choice of
$j(A,s)$ can overdetermine this ordinary (mixed) differential equation, because
the function $j(A,s)$ must be \lq separable\rq~in the sense of
Eq.\,(\ref{currenttrafoexpl2}).

An equivalent formulation of Eq.\,(\ref{currenttrafoexpl2}) can be found by eliminating all
terms and derivatives of $\alpha$. This results in the following differential equation
\begin{equation}
\displaystyle\frac{\partial^2 j}{\partial s^2}\, B_{S}
\displaystyle\frac{\partial B_{S}}
{\partial s} +
\displaystyle\frac{\partial j}{\partial s}\,\left(\left(\frac{\partial B_{S}}{\partial s}\right)^2
+ \displaystyle B_{S}\frac{\partial^2  B_{S}}{\partial s^2}\right) =0 \, ,
\label{currentrestrict}
\end{equation}
which represents a constraint for $j(A,s)$.

Because any formal integration of Eq.\,(\ref{currentrestrict}) leads automatically back to
Eq.\,(\ref{currenttrafoexpl2}), Eq.\,(\ref{currentrestrict}) is only a necessary condition,
testing or proving if any considered current density $j(A,s)$ enables the calculation of the
transformation $\alpha'(A)$ from Eq.\,(\ref{currenttrafoexpl2}).

The same integration procedure as in Eq.\,(\ref{currenttrafoexpl3}) leading to
Eq.\,(\ref{currenttrafoexpl4}) could basically also be performed for completely 2D equilibria,
which are not asymptotically 1D. The only restriction is again the integrability condition:
to guarantee the existence of an allowed transformation (to be precise, $\alpha'$ should
be an explicit function of $A$ only), the condition
\begin{eqnarray}
&& \frac{\partial\alpha'}{\partial s}=0\\
\Leftrightarrow && \displaystyle\frac{ \displaystyle\partial}{ \displaystyle\partial s}
\,\displaystyle\frac{\displaystyle\int\,\left(\exp{\int\,\displaystyle
-\frac{P_{S}\,'}{B_{S}^2} \, dA}\right)
\left(-\frac{j}{B_{S}^2}\right) \, dA + C_{0}}
{\exp{\displaystyle\left(\int\, -\frac{P_{S}\, '}{B_{S}^2} dA\right)}}=0
\label{currenttrafoexpl5}
\end{eqnarray}
must be valid. Here $C_{0}$ is an explicit function of $s$ only, $P_{S}\, '$ is an explicit
function of $A$, and $B_{S}$ and $j$ are explicit functions of $A$ and $s$. The partial
differential equation Eq.\,(\ref{currenttrafoexpl5}) is now only of first order, concerning
the $\partial/\partial s$ derivative, in contrast to the Eq.\,(\ref{currentrestrict}),
but the problem of integrability can also neither be eliminated nor solved in this way.
Conversely, the procedure in Sect.\,\ref{inversemethod} fulfills the integrability condition
Eq,\,(\ref{currenttrafoexpl5}) automatically, because it is an asymptotical 1D problem.

\begin{acknowledgements}
We thank the anonymous referee for useful comments and suggestions on the draft.
This research made use of the NASA Astrophysics Data System (ADS). D.H.N. and 
M.K. acknowledge financial support from GA\,\v{C}R under grant number 13-24782. 
M.K. also acknowledges financial support from GA\,\v{C}R under grant number 
P209/12/0103. The Astronomical Institute Ond\v{r}ejov is supported by the 
project RVO:67985815.
\end{acknowledgements}

\bibliographystyle{aa} 
\bibliography{flowmhd} 

\end{document}